\documentclass[12pt,preprint]{aastex}
\usepackage{natbib} 
\newcommand{\msun}{$M_{\odot}$}
\newcommand{\lsun}{$L_{\odot}$}

\newcommand{\Mbh}{{M_{\rm BH}}}

\newcommand{\Mdot}{\dot{M}_{\rm acc}}

\newcommand{\LFIR}{L_{\rm FIR}}

\shorttitle{Molecular Gas in IR QSOs}
\shortauthors{Xia et al.}

\begin{document}

\title{Molecular Gas in Infrared Ultraluminous QSO Hosts}
\author{X.Y. Xia\altaffilmark{1},
Y. Gao\altaffilmark{2, 3},
C.-N. Hao\altaffilmark{1},
Q. H. Tan\altaffilmark{2, 3, 4},
S. Mao\altaffilmark{5,6},
A. Omont\altaffilmark{7},
B. O. Flaquer\altaffilmark{8},
S. Leon\altaffilmark{8, 9},
P. Cox\altaffilmark{10}
}

\altaffiltext{1}{Tianjin Astrophysics Center, Tianjin Normal University, Tianjin 300387, China; xyxia@bao.ac.cn}
\altaffiltext{2}{Purple Mountain Observatory, Chinese Academy of Sciences, 2 West Beijing Road, Nanjing 210008, China}
\altaffiltext{3}{Key Laboratory of Radio Astronomy, Chinese Academy of Sciences, Nanjing 210008, China}
\altaffiltext{4}{Graduate School of Chinese Academy of Sciences, Beijing 100039, China}
\altaffiltext{5}{National Astronomical Observatories of China, 20A Datun Road, Chaoyang District, Beijing 100012, China}
\altaffiltext{6}{Jodrell Bank Centre for Astrophysics, University of Manchester, Alan Turing Building, Manchester M13 9PL, UK}
\altaffiltext{7}{Institut d'Astrophysique de Paris, UMR7095, UPMC and CNRS, 98bis boulevard Arago, F-75014, Paris, France}
\altaffiltext{8}{Instituto de Radioastronom\'{i}a Milim\'etrica (IRAM), Avenida Divina Pastora 7, N\'{u}cleo Central, 18012 Granada, Spain}
\altaffiltext{9}{Joint Alma Observatory/ESO, Av. El Golf 40, Piso 18, Las Condes, Santiago, Chile}
\altaffiltext{10}{Institut de Radio Astronomie Millimetrique (IRAM), St. Martin d'Heres, France}

\begin{abstract} We report CO detections in 17 out of 19 infrared ultraluminous
QSO (IR QSO) hosts observed with the IRAM 30m telescope. The cold molecular gas
reservoir in these objects is in a range of 0.2--2.1$\times 10^{10}M_\odot$
(adopting a CO-to-${\rm H_2}$ conversion factor $\alpha_{\rm CO}=0.8 M_\odot\,
{\rm (K~km\,s^{-1}~pc^2)^{-1}}$). We find that the molecular gas properties of
IR QSOs, such as the molecular gas mass, star formation efficiency ($L_{\rm
FIR}/L^\prime_{\rm CO}$) and the CO (1-0) line widths, are indistinguishable
from those of local ultraluminous infrared galaxies (ULIRGs).  A comparison of
low- and high-redshift CO detected QSOs reveals a tight correlation between
L$_{\rm FIR}$ and $L^\prime_{\rm CO(1-0)}$ for all QSOs. This suggests that,
similar to ULIRGs, the far-infrared emissions of all QSOs are mainly from dust
heated by star formation rather than by active galactic nuclei (AGNs),
confirming similar findings from mid-infrared spectroscopic observations by
{\it Spitzer}.  A correlation between the AGN-associated bolometric luminosities
and the CO line luminosities suggests that star formation and AGNs draw from the
same reservoir of gas and there is a link between star formation on $\sim$ kpc
scale and the central black hole accretion process on much smaller scales.
\end{abstract}


\section{INTRODUCTION}

It has become increasingly clear that the growth of central supermassive black
holes (SMBHs) and their host spheroids are closely related since the mass of
the SMBH correlates well with the properties of the hot stellar component of
the host galaxy, e.g. the galactic velocity dispersion
($\Mbh$-$\sigma_\star$), the luminosity/mass of the spheroidal component of the
host galaxy (e.g.  Magorrian et al. 1998; Ferrarese \& Merritt 2000; Tremaine
et al. 2002; H\"aring \& Rix 2004). Hydrodynamical simulations on gas-rich
galaxy mergers, incorporating star formation, SMBH growth and feedback from
both supernovae and SMBH, provide a plausible picture of how galaxy formation
and the growth of SMBHs are related to each other (e.g.  Hopkins et al. 2006).
However, much of the gas physics in numerical simulations, in particular how
the star formation and central active galactic nucleus (AGN) activities feed
energy and momentum back into the gas, is uncertain. Such feedbacks may heat up
the cold gas and drive a powerful galactic wind or outflow, which may limit or even
terminate further star formation and black hole growth in galaxies.

Hao et al. (2005, hereafter H05) studied a sample of type I active galactic
nuclei (AGNs) selected from local ultraluminous infrared galaxies (ULIRGs, the
$8-1000\mu$m integrated infrared luminosities are larger than $10^{12}$\lsun) 
and referred these as IR QSOs. By comparing IR QSOs with optically selected PG
QSOs and narrow-line Seyfert 1s (NLS1s) in the relation of the IR luminosities
versus the optical luminosities, they find that IR QSOs show mid-IR and far-IR (especially
far-IR, hereafter FIR) excess. The FIR luminosities for most IR QSOs are larger
than half of their central AGN bolometric luminosities. In contrast, the median ratio of the FIR to central AGN bolometric
luminosities for local classical QSOs is four times smaller than that for IR QSOs.
Comparisons of the FIR spectral index of IR
QSOs with those of PG QSOs indicate that the FIR excess of IR QSOs relative to
the PG QSOs is suggestive of massive starbursts.  These IR QSOs not only have
massive starbursts occurring in their host galaxies, but their optical
spectroscopic and X-ray properties also exhibit characteristics of young,
growing QSOs with high accretion rates to their central black holes (Zheng et
al. 2002; H05).  Moreover, images of these IR QSOs show that most are
undergoing clear tidal interactions or major mergers (Canalizo \& Stockton
2001; see also Fig. 1). The optical spectra show obvious blueshifts in
permitted emission lines in some IR QSOs, indicating outflows.  More recent
IRAM Plateau de Bure Interferometer (PdBI) and Herschel observations of Mrk
231, the closest IR QSO, reveal clear outflows with velocity larger than 1000
km\,s$^{-1}$, as inferred from broad wings of the CO line or OH P-Cygni profile
(Feruglio et al. 2010; Fischer et al. 2010) and the coexistence of active SMBH
accretion and strong starburst (van der Werf et al. 2010).  The mid-IR spectra
of IR QSOs from {\it Spitzer}/Infrared Spectrograph (IRS)  also show that the
typical slope of spectral energy distribution (SED) for IR QSOs is between
those of classical PG QSOs and ULIRGs (Cao et al. 2008). All the properties of
IR QSOs are consistent with their being at an important transition stage from
mergers with massive starbursts (ULIRGs) to dust-enshrouded QSOs and later
luminous QSOs, and eventually early-type galaxies (Sanders et al. 1988a; 
Sanders \& Mirabel 1996; Tacconi et al.2002).  Studies of these objects may 
thus provide significant insights into the understanding of how the growths 
of spheroids and SMBHs are inter-connected and how the $\Mbh$-$M_\star$ relation 
evolves as a function of redshift.
                                                                                                                                                
The conclusion that the FIR excess in IR QSOs is primarily due to starbursts
requires a large reservoir of cold molecular gas in IR QSO hosts.  In fact, up to
now only 6 local IR QSOs (I Zw I , Mrk 231, Mrk 1014, 3C 48, IRAS 1613+658 and PG
1700+518) have been observed (Solomon et al. 1997; Schinnerer, Eckart \&
Tacconi 1998; Krips et al.  2005; Evans et al. 2001, 2009). The molecular gas
masses of these IR QSO hosts are few times  $10^{9}$ to $10^{10}$\msun, assuming the
same conversion factor from the CO luminosity to $H_{2}$ mass as that for
ULIRGs, i.e.  $\alpha_{\rm CO}=0.8 M_\odot\, {\rm (K\,km\,s^{-1}~pc^2)^{-1}}$
(Downes \& Solomon 1998). In contrast, the molecular gas masses of CO detected
classical QSO hosts are at most a few times $10^{9}$ \msun\ (Evans et al. 2001; 
Scoville et al. 2003). The larger molecular gas reservoirs in these IR QSO hosts
mean that there are rich of fuel to sustain the star formation and may
support the speculation by Hao et al.  (H05) that the FIR excess in IR QSOs is
mainly due to starbursts. However,  CO observations for a larger sample of IR
QSOs are needed to statistically confirm this scenario.

As we will show later, in the IR QSO phase the SMBH and spheroid are both
growing rapidly.  Two questions naturally arise: What is the timescale for the
consumption of the molecular gas to sustain the massive starbursts?  Can the
local $\Mbh$-$M_\star$ relation be established/maintained in this phase?
Such a spheroid and central SMBH co-evolution process is expected to be more
important at high redshift according to the downsizing galaxy formation
scenario (Cowie et al. 1996). While a few tens of high-redshift QSOs have been
detected in CO (Solomon \& Vanden Bout 2005; Coppin et al. 2008; Wang et al.
2010; Riechers 2011a; Polletta et al. 2011), surprisingly only half a dozen
local IR QSOs have been observed before this work.  Clearly a larger local IR
QSO sample detected in CO is desirable for comparisons with high-redshift CO
detected QSOs. For all the above reasons, we have initiated a program to observe a larger
sample of IR QSOs with the IRAM 30m telescope in order to systematically
investigate the molecular gas properties of these objects. 

This paper presents the results of our CO observations of 19 IR QSOs with
the IRAM 30m. The paper is structured as follows. In Sections 2 and 3, we describe our
sample and outline the observations and data reduction respectively. We present
our main results and discussion in Section 4, and a summary in Section 5.
Throughout this paper we adopt a cosmology with a matter density parameter
$\Omega_{\rm m}=0.3$, a cosmological constant $\Omega_{\rm \Lambda}=0.7$ and a
Hubble constant of $H_{\rm 0}=70\,{\rm km \, s^{-1} Mpc^{-1}}$.

\section{SAMPLE}

\begin{figure}
\centering
\includegraphics[width=1.0\textwidth]{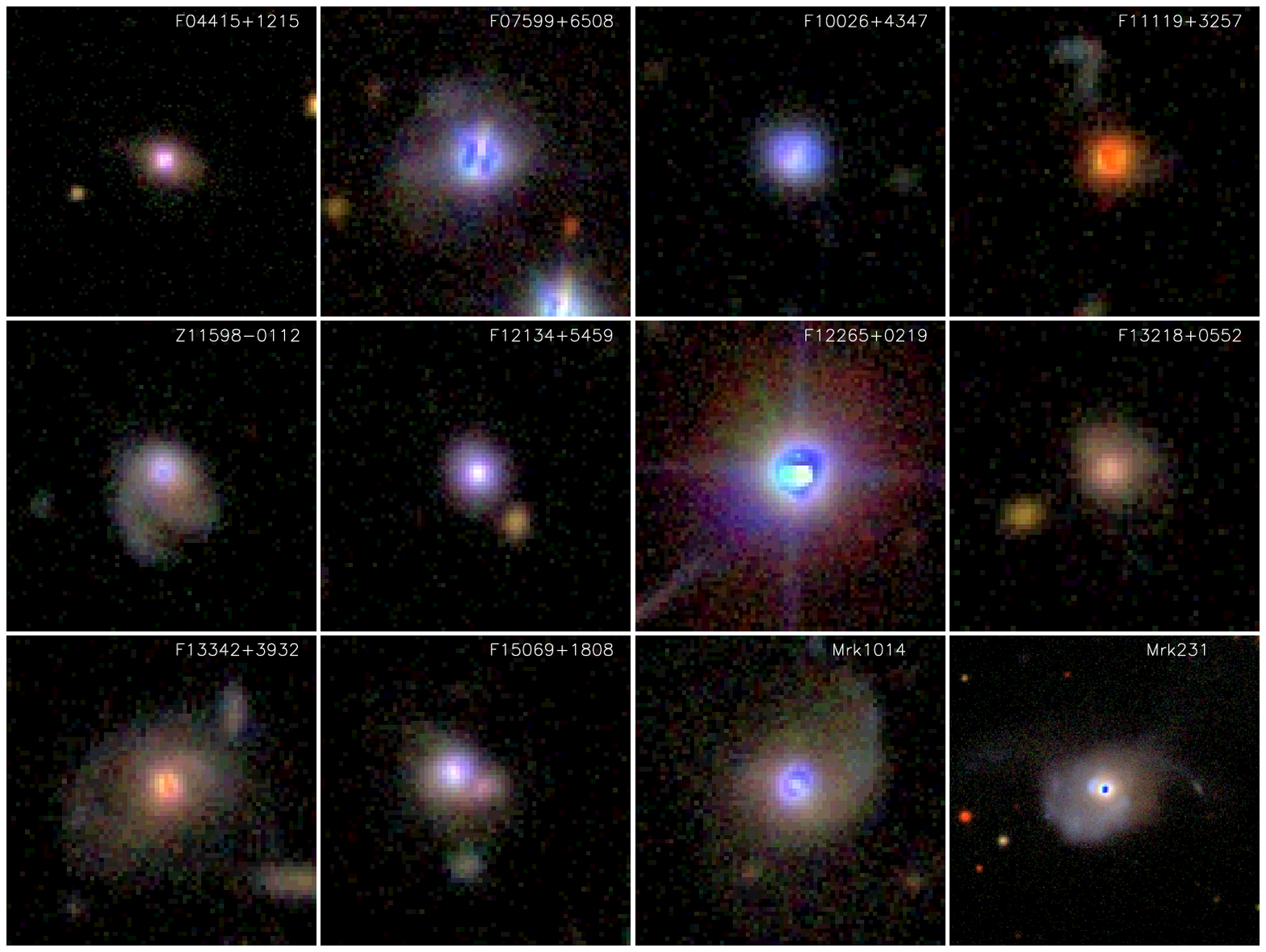}
\caption{True color images of 12 IR QSOs (out of a total of 25), constructed from SDSS g, r, and i images using color-preserving nonlinear stretches (Lupton et al. 2004). 
The box size of each image is 80\,kpc.
}
\label{irqso.ima-fig1.eps}
\end{figure}

All of our 19 newly CO observed IR QSOs are selected from the IR QSO sample of
H05 and have redshift $z< 0.2$ except for $F13218+0552$ (z=0.205), $F18216+6419$ (z=0.297)
and $F02054+0835$ (z=0.345). This IR QSO
sample (H05) was compiled from the ULIRGs in the QDOT redshift survey (Lawrence
et al. 1999), the 1$\,$Jy ULIRG survey (Kim et al. 1998), and the
cross-correlation of the IRAS Point Source Catalog (PSC) with the $ROSAT$
All-sky Survey Catalog. We also include six previously published CO detections
of IR QSOs. In total, our sample consists of 25 IR QSOs.  As the fraction of IR
QSOs in ULIRGs is less than 10\% and the PSC IRAS redshift survey (Saunders et
al. 2000) galaxy sample includes about 1000 ULIRGs, so the total number of IR
QSOs expected in the northern sky is $\lesssim$ 100.  Our IR QSO sample
includes about a quarter of the total,  and thus should be a representative IR
QSO sample in the local universe (see Zheng et al. 2002 for more detailed
descriptions).  Fig.~\ref{irqso.ima-fig1.eps} shows true color images of 12 IR
QSOs (out of 25) cross-identified from DR7 of the Sloan Digital Sky Survey
(SDSS).  Most of these show tidal features and may be in the final merging
stage.

For comparisons, we also compiled several samples with CO
detections, including local ULIRGs, low-redshift classical QSOs and
high-redshift (sub)millimeter (mm) luminous QSOs. Notice that we correct all the published
results to our adopted cosmology.  In the following, we briefly summarize these
samples.  
\begin{enumerate} 
\item[(1)] A ULIRG sample taken from Solomon et
al. (1997) consists of 36 objects in the redshift range of $z=0.03$ to 0.27.
The CO observations were made with the IRAM 30m telescope. There are more 
ULIRGs with CO observations (e.g., Chung et al. 2009; Mirabel et al. 1990), but here we
restrain ourselves to systems with IRAM 30m data.
\item[(2)] A local classical QSO sample with CO detections is compiled from 11 PG QSOs
(Evans et al. 2001; Scoville et al. 2003) and 16 HE QSOs (Bertram et
al. 2007), after excluding several IR QSOs. Evans et al. (2001) selected their sample
by IR excess (L(8-1000$\mu$m)/L(0.1-1.0$\mu$m) $> 0.36$) in the
redshift range of $0.04<z<0.17$, while Scoville et al. (2003) selected their PG
QSO sample with criteria of $z<0.1$ and $M_{\rm B}<-23$ mag.  The HE QSO
sample is a nearby low-luminosity QSO sample with redshift $z<0.06$, 
drawn from the Hamburg-ESO survey for bright UV-excess QSOs and observed
also by the IRAM 30m telescope (Bertram et al. 2007). 
\item[(3)] The (sub)mm luminous high-redshift QSO sample of 29 objects is taken from Wang et al.
(2010). They observed six (sub)mm luminous QSOs with z$\sim$6 
by the IRAM PdBI; for each object, either the CO(6-5) or (5-4) line emission has been detected. 
They also included two  QSOs at z$\sim$6 previously detected in CO and another 21 
high-redshift CO detected QSOs in the range of $\rm
1.4\leq z\leq5$ from the literature.  Notice that
this CO detected, high-redshift (sub)mm luminous QSO sample includes nearly all such objects up to 2010.  
\end{enumerate}

\section{OBSERVATIONS, DATA REDUCTION and PARAMETER ESTIMATION}

The CO observations were carried out with the IRAM 30m telescope on Pico
Veleta near Granada, Spain. Although the first 
of such attempts was conducted remotely from the IRAM headquarter in Grenoble 
in July 2007, the weather was mostly bad then and essentially little 
useful data were obtained. The data used here were mainly collected 
during the summer and autumn of 2008. Both the 3mm and 1mm 
dual-polarization receivers were
used together with the 512$\times$1 MHz and 256$\times$4 MHz filter banks to
simultaneously obtain both CO (1$-$0) and CO (2$-$1) emission lines.
The weather was mostly excellent during these observing runs, but 
some CO (2$-$1) lines were still not observed for a few targets since 
the line frequencies were redshifted out of the tuning range of 1mm receivers.
All observations were done with a wobbler switching mode with a throw rate of 0.5 Hz and a
wobbler throw of $60\arcsec$ in azimuth. System temperatures were on average
$T_{\rm sys}$\,$\approx$\,150\,K and 450\,K at $\sim$ 105\,GHz and $\sim$ 210\,GHz,
respectively. The total usable on-source
integration time is  $\sim$50 hours from the 2008 runs.  Table 1 summarizes the details of CO observations
for all 19 sources.

We reduced the data using the IRAM software CLASS/GILDAS. First, we discarded
a few bad scans with strongly distorted baselines, fixed some bad channels
before summing up all scans to look for the likely detected CO line emission
windows. We then subtracted linear baselines in each of the remaining scans
with the adopted velocity windows before summing up to obtain the final
spectra that are further subtracted with linear baselines.  In order to
increase the signal-to-noise ratio, all the spectra were Hanning smoothed to
velocity widths of $>20$\,km s$^{-1}$.  The amplitudes of each spectrum were
converted to main beam temperatures, $T_{\mathrm{mb}}$, by multiplying
$T_{\mathrm{A}}^{\ast}$ by the ratio between the forward efficiency and main
beam efficiency $F_\mathrm{eff}/B_\mathrm{eff}$. The main beam efficiency
$B_{\mathrm{eff}}=0.75$ and the forward efficiency $F_{\mathrm{eff}}=0.95$ were
applied to $T_{\mathrm{A}}^{\ast}$ at $\sim$ 105 GHz and
$B_{\mathrm{eff}}=0.57$ and $F_{\mathrm{eff}}=0.91$ to $T_{\mathrm{A}}^{\ast}$
at $\sim$ 210 GHz, respectively. For point sources, flux densities were
obtained using the conversion factor $S_{\nu}/T_\mathrm{mb}=4.95$ Jy/K for the IRAM 30m telescope.

The CO line luminosities $L^\prime_{\rm CO}$ for IR QSOs were calculated following Solomon, Downes, \& Radford (1992) 
\begin{equation}
{L^\prime_{\rm CO}=23.5 \;\Omega_{S\star B}\;D_{\rm L}^2 \; I_{\rm CO}\; (1+z)^{-3}\;\rm \left[K\; km\; s^{-1}\; pc^2\right] },
\end{equation}
where $I_{\rm CO}$ is the velocity-integrated line intensity in units of K km s$^{-1}$, ${ D_{\rm L}}$ is
the luminosity distance measured in Mpc, and ${\rm \Omega_{S\star B}}$ is the solid angle of the source
convolved with the telescope beam in square arc-seconds, i.e.
${\rm \Omega_{S\star B}}$=$\pi\over{4 \ln2}$$ (\theta_{S}^2+\theta_{B}^2)$, where $\theta_{B}=24\arcsec$ at 105 Ghz.

The CO luminosities for PG and HE QSOs were estimated using the same method 
and were provided in the respective papers described in \S 2. We simply adapted
them to our adopted cosmology. For the high-redshift QSO sample collected by Wang et al. (2010), a different approach
was applied since most of these objects were detected at high orders
($J>5$) of CO transitions. Wang et al. (2010) calculated the $L^\prime_{\rm CO(1-0)}$ values for all high-z sample QSOs
under the non-thermalized excitation assumption and adopted the line ratios of CO (6$-$5)/CO (1$-$0) and CO (5$-$4)/CO (1$-$0)
as 0.78 and 0.88, respectively. Since only the CO ground state
transition CO (1$-$0) traces molecular gas at low excitation and provides a direct
estimate of the total cold molecular gas mass
(Riechers et al. 2007, 2011b; Gao \& Solomon 2004a),
we will use either the observed and/or converted CO (1$-$0) properties for all QSOs at
different redshifts in this work. 

Apart from CO line luminosities, FIR luminosities are needed as well
in this work.  The FIR luminosities
(L$_{\rm FIR}$, the luminosity in the range of 42.5-122.5$\mu$m) were
calculated based on the flux densities at 60$\mu$m and 100$\mu$m from the IRAS Faint
Source Catalog for local IR QSOs, ULIRGs and PG$+$HE QSOs, following Helou et
al. (1988). For high-redshift (sub)mm loud QSOs, their FIR luminosities were
provided by Wang et al. (2010; private communication); they  derived L$_{\rm FIR}$
based on (sub)mm continuum measurements under the assumption that the
rest-frame FIR SED can be described by a grey-body
spectrum with a dust temperature of 47\,K and emissivity index 
of 1.6 (Beelen et al. 2006) except for J1148+5251 and J0927+2001. For these two
objects, Wang et al. (2010) used a fitted temperature of 56\,K  and 46\,K respectively.

We also need the AGN-associated bolometric luminosities $L_{\rm AGN}$ for our 
QSO samples (hereafter we use $L_{\rm AGN}$ to denote the bolometric luminosities from AGN for all QSOs).
For local PG QSOs and IR QSOs, they were estimated from the extinction
corrected continuum emission at 5100\AA\ and adopting a bolometric correction
factor of 9, i.e., $L_{\rm AGN}\approx9\lambda L_{\lambda}(5100){\rm
\mbox{\AA}}$ (H05).  For HE QSOs, the central AGN bolometric luminosities $L_{\rm AGN}$ were derived from
their ${\rm B}_{\rm J}$ magnitudes with a bolometric correction factor of 9.74
(Vestergaard et al. 2004).  For  the 11 high-redshift QSOs in Hao et al.
(2008), $L_{\rm AGN}$ were taken from that paper directly,
calculated in the same way as for HE QSOs.  For the eight z $\sim 6$ QSOs, the
absolute AB magnitudes at rest frame 1450\AA\ derived by Wang et al.  (2010)
were converted into the B band magnitudes according to Schmidt et al. (1995;
see also Fan et al.  2001), and then a bolometric correction factor of 9.74 was
employed to convert the B-band magnitudes into $L_{\rm AGN}$.  For
the ten high-redshift QSOs neither in Hao et al. (2008) nor in Wang et al.
(2010), we computed their $L_{\rm AGN}$ in the same way as Hao et al. (2008).

\section{RESULTS AND DISCUSSIONS}

\subsection{CO Emissions from IR QSO hosts}

\begin{figure}
\centering
\includegraphics[width=1.\textwidth,angle=0]{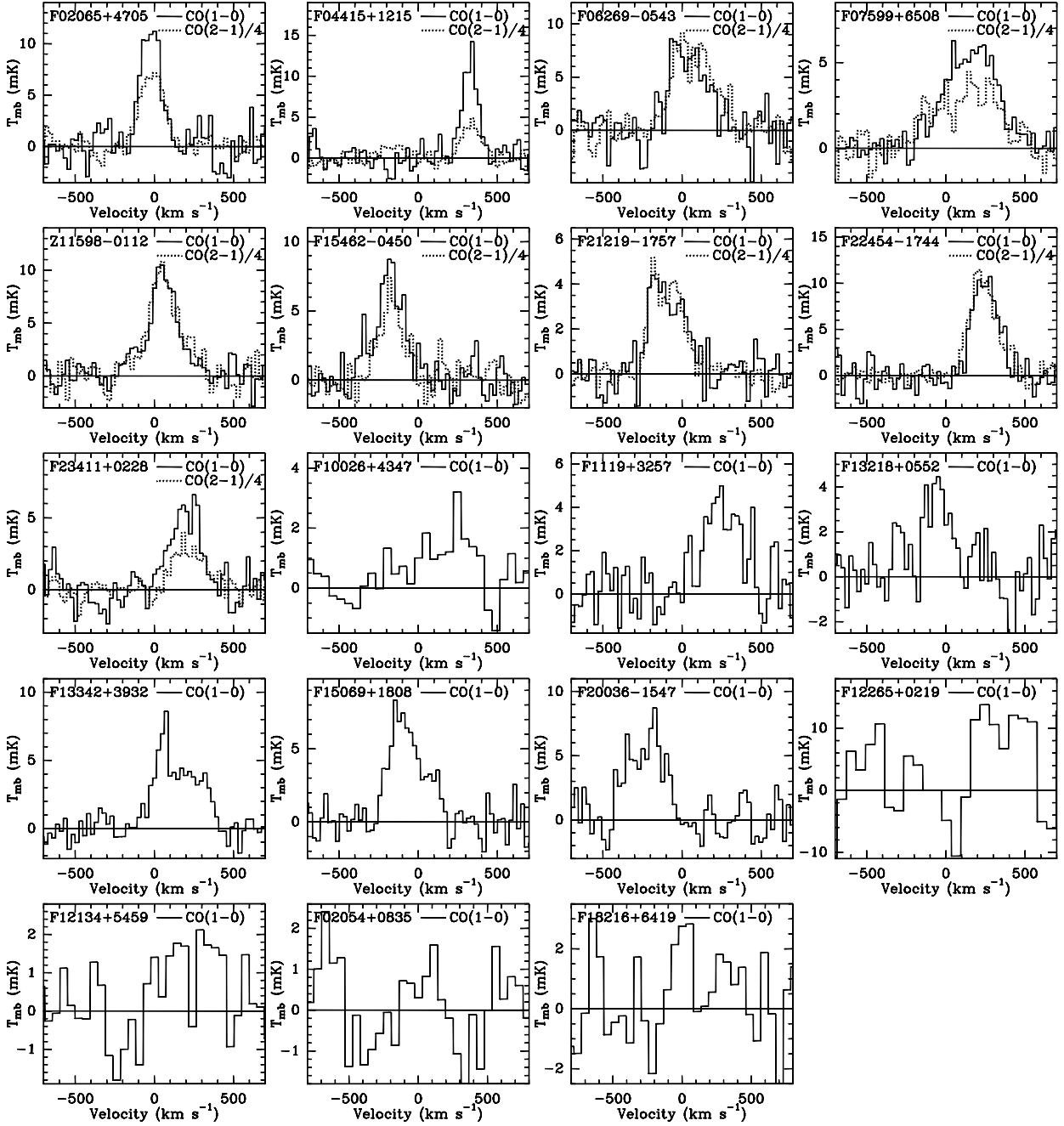}
\caption{Spectra of CO (1-0) (solid lines) and CO (2-1) (dotted lines, divided by a factor of 4 for 9 targets) emission for our IR QSO sample. The spectra are smoothed to 23 km s$^{-1}$ channel width for display. Only marginal detections ($3\sigma$) in F12265 and F12134, and non-detection in F02054 and F18216; for these objects, the spectra are smoothed to 46 km s$^{-1}$. The main-beam temperature can be converted to the flux density using S/T$_{\rm mb}$=4.95 Jy/K. }
\label{spectra-all-fig2.eps}
\end{figure}

17 out of the 19 new IR QSO hosts have been detected in CO (1$-$0), while nine of
them have also been observed and all detected in  CO (2$-$1). All the
detections except for F12265+0219 and F12134+5459 (which were marginally
detected at 3$\sigma$) have signal-to-noise ratio (S/N) greater than 5$\sigma$.
Fig.~\ref{spectra-all-fig2.eps} presents the CO (1$-$0) and CO (2$-$1) spectra
for all IR QSOs. Those targets without CO (2$-$1) lines are simply redshifted
out of tunable range of the 1mm receivers and thus cannot be observed.  We also
calculated the 3$\sigma$ upper limits (about $10^{10}\,{\rm K\,km\, s^{-1}\,
pc^2}$) for the two non-detections by assuming a line width (Full Width at Zero
Intensity, hereafter FWZI) of 600\,km s$^{-1}$. The reasons for non-detections
(for  IRAS F18216+6419 and IRAS F02054+0835) may be that the integration time
is insufficient because these two objects have the highest redshifts in our
sample. 

The CO (1$-$0) and CO (2$-$1) line widths (FWZI), integrated line luminosity,
FIR luminosity and L$_{\rm FIR}$/L$^{'}_{\rm CO(1-0)}$ for all our
sample objects, including 6 IR QSOs from the literature, are listed in Table 2.
From this table, we can see that the CO luminosities ($L^\prime_{\rm CO}$) for
the 23 CO detected IR QSO hosts (17 of which are from our sample) are several times
10$^9$ to a few times 10$^{10}$\,K\, km s$^{-1}$ pc$^2$, comparable to those for
ULIRGs with the largest molecular gas reservoirs in the local universe. The upper
limits of the two non-detections are also consistent with this
mass range.

The CO (2$-$1)/CO (1$-$0) line luminosity ratio ($r_{21}$) ranges from 0.4 to
1.2 with mean value of $\sim 0.8$ in nine CO (2$-$1) detected IR QSOs.  Four IR
QSOs out of nine are found to have $r_{21}\geq 1$ with uncertainty of $\sim 15\%$, suggesting that the
molecular gas is essentially thermalized and has a point-like distribution
compared to the $12\arcsec$ CO(2$-$1) beam (corresponding to $24$\,kpc for the
median redshift of CO (2$-$1) detected IR QSOs, $z\sim 0.11$).  Three IR QSOs
have $r_{21} \sim 0.7$ with uncertainty of $\sim 12\%$, a typical average value for ULIRGs (Radford, Downes \&
Solomon 1991), which implies they, similar to ULIRGs, have modest sub-thermal
molecular gas properties. Furthermore, if they are indeed sub-thermally excited
with such line ratios, then these sources again have point-like molecular gas
distribution compared to the $12\arcsec$ beam.  Only 2  IR QSOs have really low
$r_{21}$ ratios ($\sim 0.5$) with uncertainty of $\sim 10\%$, which may be due to their sub-thermalized
molecular gas properties and/or their extended spatial structure (Garay,
Mardones \& Mirabel 1993) of quite a few arc-seconds. From our limited CO
(2$-$1) data, we infer that the molecular gas distributions in most (7 out of
9) IR QSOs are probably concentrated in compact nuclear regions.  However,
further high spatial resolution observations, e.g., Aravena et al.  (2011), are
needed to reveal the spatial distribution and kinematics of the molecular gas
in IR QSOs.

\subsection{Comparison with Local ULIRGs}

\begin{figure}
\centering
\includegraphics[width=1.0\textwidth]{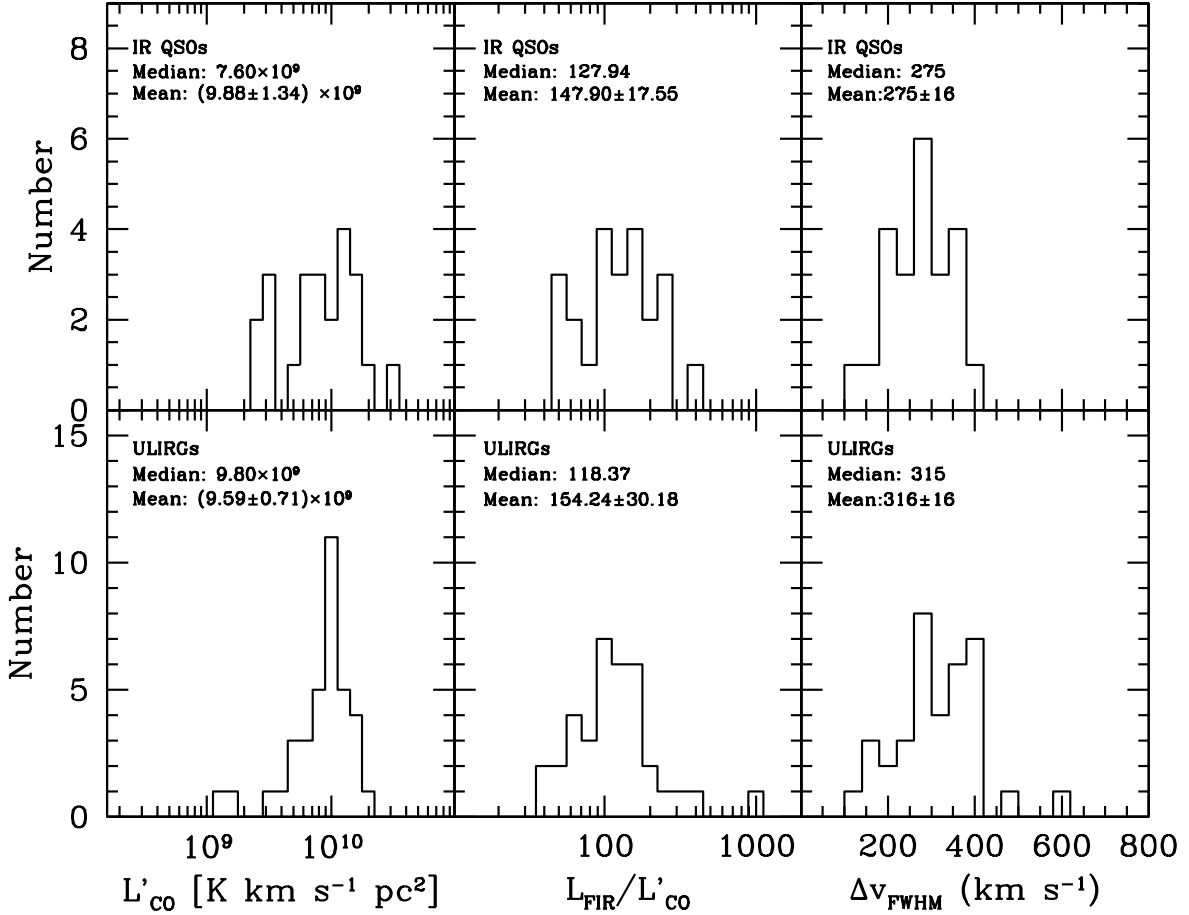}
\caption{Histograms of CO (1-0)
luminosity L$^{'}_{\rm CO(1-0)}$, star formation efficiency 
(ratio of far-infrared luminosity to CO luminosity) $L_{\rm FIR}/L^\prime_{\rm CO(1-0)}$
 and CO line width (FWHM) for IR QSO hosts (upper panels) and ULIRGs (lower panels). The median and mean values are indicated in the top-left corner of each panel.
}
\label{irqsoulirgs.his-fig3.eps}
\end{figure}

Fig.~\ref{irqsoulirgs.his-fig3.eps} compares the CO properties of IR QSO hosts
(upper panels) with those of ULIRGs (lower panels). Specifically, the
histograms of the CO luminosity (L$^{'}_{\rm CO(1-0)}$) (left panels), the FIR
to CO luminosity ratio $L_{\rm FIR}/L^\prime_{\rm CO}$ (commonly used as an
indicator of the star formation efficiency [SFE], middle panels) and CO (1$-$0)
line width (FWHM, right panels) are compared. Median and mean values are
labelled in the upper left corner of the panels.

From Fig.~\ref{irqsoulirgs.his-fig3.eps} we can see that the median value of
L$^{'}_{\rm CO(1-0)}$ for IR QSO hosts is similar to that of ULIRGs.  Hence, the
molecular gas in IR QSO hosts is as abundant as in ULIRGs which can easily provide
the fuel needed for their ongoing starbursts.  In addition, the median values
of $L_{\rm FIR}/L^\prime_{\rm CO}$ (SFE) and CO line width for IR QSO hosts are also
similar to those of ULIRGs at the $3\sigma$ level.  IR QSO hosts appear to have
similar SFRs, velocity spread in CO, and SFEs as ULIRGs,  likely though ULIRGs
have evolved into the QSO episode manifested as IR QSOs. 

It is widely accepted that the FIR emission of ULIRGs is mainly from star
formation (e.g. Yun et al. 2001; Gao \& Solomon 2004b) 
and most molecular gas in ULIRGs is distributed in centrally rotating
nuclear disks or rings as revealed by CO images (Downes \& Solomon 1998). The
similarities of IR QSO hosts and ULIRGs in their CO properties suggest that their
FIR luminosities are both dominated by dust heated by starbursts and even their
cold molecular gas  may have similar spatial distributions. In fact, the
highest resolution CO interferometer observations for two IR QSOs, Mrk 231 and
I Zw I show nuclear disks and rings.  For Mrk 231, there is an inner disk of
diameter $1.2\arcsec$ (radius $\sim$ 0.5 kpc) and an outer disk with diameter
of  $3\arcsec$ (radius $\sim$ 1.25 kpc,  Downes \& Solomon 1998).  For I Zw I,
there is a ring-like structure of radius 1.2 kpc (Staguhn et al. 2004). Our
available line ratios of $r_{21}$ are also consistent with the CO emission in
IR QSOs being mostly concentrated in nuclear regions.  If most IR QSO hosts indeed
have similar CO properties to those of ULIRGs, it follows that the CO-to-${\rm
H_2}$ conversion factor from $L^\prime_{\rm CO}$ to cold molecular gas mass for
IR QSOs should be the same for ULIRGs, i.e. $\alpha_{\rm CO}=0.8 M_\odot~ {\rm
(K~km\,s^{-1}~pc^2)^{-1}}$.  Therefore, the cold molecular gas properties, such
as cold molecular gas mass, SFE, and possibly the spatial distribution in IR
QSOs, are indistinguishable from those of ULIRGs.

Taking into account the properties of IR QSOs in different wavebands, 
such as the optical morphology, extreme \ion{Fe}{2} emissions in their optical spectra
(Zheng et al. 2002), FIR excess emissions compared with classical 
QSOs (H05) and similar 6.2$\mu$m PAH and [NeII] 12.81$\mu$m emission
strengths in IR QSO and ULIRGs (Cao et al. 2008), there is little doubt that
there are ongoing massive starbursts in IR QSO hosts, as in ULIRGs. The timescale of 
starbursts in IR QSOs is about a few times $10^{7}$\,yr, as inferred
from $M_{\rm gas}/{\rm SFR}$ by assuming that the FIR emission is
predominately from star formation (see Table 3).

\subsection{Comparison with Local and High-redshift CO-detected QSOs}

\begin{figure} 
\centering 
\includegraphics[width=1.0\textwidth]{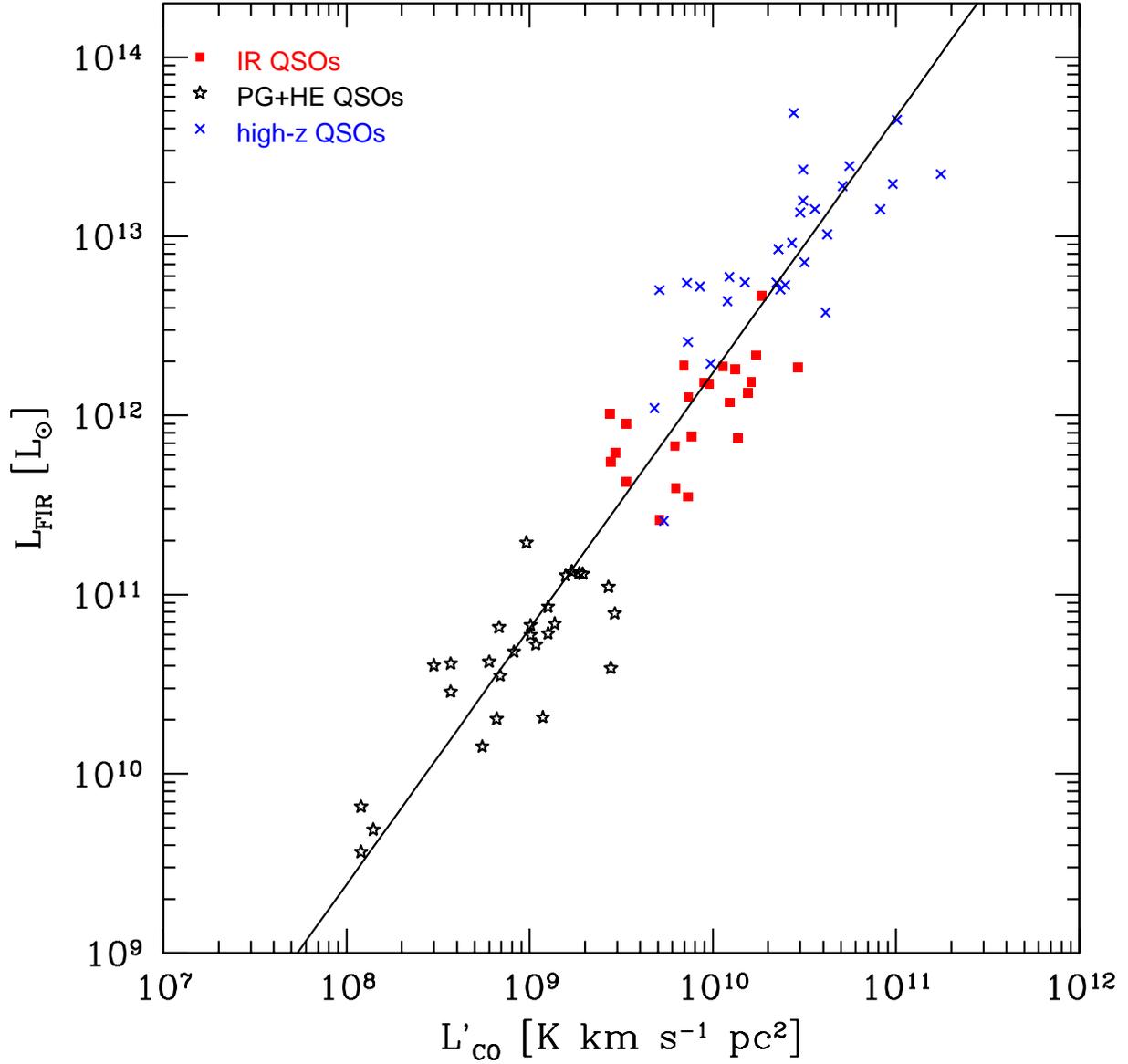}
\caption{L$_{\rm FIR}$ vs. $L^\prime_{\rm CO(1-0)}$ for almost all CO detected IR QSOs (red
filled squares), PG$+$HE QSOs (black open pentagrams) and high-redshift QSOs (blue
crosses). The samples are described in \S2.  A least-square bisector best-fit
line obtained for all objects is shown; the best-fit power-law slope is $1.4\pm0.1$.  }
\label{Lfir.LCOprime-fig4.eps} 
\end{figure}

Fig.~\ref{Lfir.LCOprime-fig4.eps} presents the L$_{\rm FIR}$ versus
$L^\prime_{\rm CO(1-0)}$ relation for almost all local and high-redshift QSOs
with CO detections up to 2010. Throughout this paper, the red filled squares
represent the local IR QSOs, black open pentagrams denote the local PG and
HE QSOs and blue crosses represent the high-redshift QSOs.  It is clear
from Fig.~\ref{Lfir.LCOprime-fig4.eps} that there is a tight correlation
between L$_{\rm FIR}$ and $L^\prime_{\rm CO(1-0)}$, spanning four and three
orders of magnitude in L$_{\rm FIR}$ and $L^\prime_{\rm CO(1-0)}$,
respectively.  The Spearman Rank-order (S-R) correlation analysis gives a
coefficient of 0.94 with significance of $>$ 99.99$\%$.  Such a strong
correlation between L$_{\rm FIR}$ and $L^\prime_{\rm CO(1-0)}$ for all CO
detected QSOs seems to suggest that the bulk of FIR emissions from all QSOs are
from the same origin (star formation), since similar correlations are well
established for star-forming galaxies at both low and high redshifts (e.g.
Solomon \& Vanden Bout 2005; Riechers 2011a). 
The best-fit power-law slope between  $L_{\rm FIR}$ and $L^\prime_{\rm CO(1-0)}$ is
approximately 1.4 regardless of the sample QSOs included. This is different from 
star-forming galaxies where the best-fit slope varies from 1 to 1.7 depending on
how many ULIRGs are included (Gao \& Solomon 2004b). 

\begin{figure}
\centering
\includegraphics[width=1.0\textwidth]{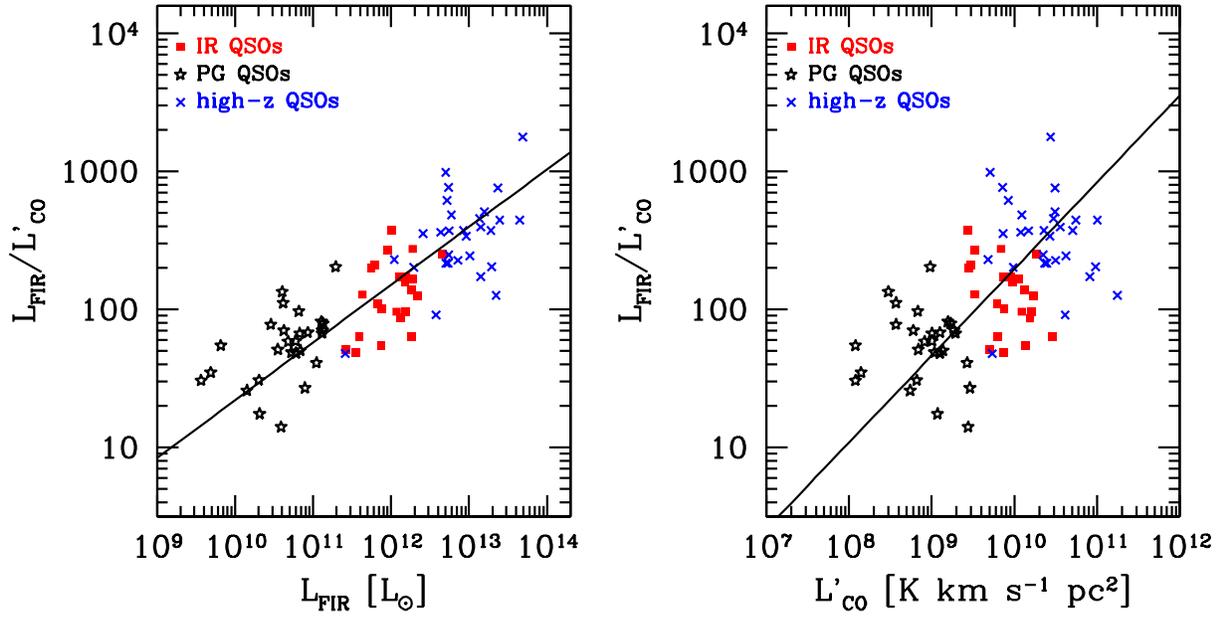}
\caption{The star formation efficiency L$_{\rm FIR}$/L$^{'}_{\rm CO(1-0)}$ vs. L$_{\rm FIR}$ (left panel)
and L$_{\rm FIR}$/L$^{'}_{\rm CO(1-0)}$ vs. $L^\prime_{\rm CO(1-0)}$
(right panel) for three samples of QSOs (see \S4.2). A Spearman Rank-order correlation analysis gives 
significances of $>$ 99.99$\%$ for both correlations. Notice that the vertical axis is independent of distance.
}
\label{LfirLCOprimerat.Lfir.LCOprime-fig5.eps}
\end{figure}

The nonlinear correlation between L$_{\rm FIR}$ and $L^\prime_{\rm CO(1-0)}$ for local star-forming
galaxies, luminous starbursts, ULIRGs
and high-redshift CO detected objects has been widely discussed and is likely due
to different dense molecular gas fractions in different classes of objects
(Gao \& Solomon 2004b; Daddi et al. 2010). Our super-linear 
power-law slope (1.4) is consistent with their conclusion.
Furthermore, Fig.~\ref{LfirLCOprimerat.Lfir.LCOprime-fig5.eps} shows
the luminosity ratio L$_{\rm FIR}$/L$^{'}_{\rm CO(1-0)}$ as functions of
FIR luminosities L$_{\rm FIR}$ (left panel) and CO (1-0) luminosities
$L^\prime_{\rm CO(1-0)}$ (right panel).  It is clear from
Fig.~\ref{LfirLCOprimerat.Lfir.LCOprime-fig5.eps} that there are tight
correlations between L$_{\rm FIR}$/L$^{'}_{\rm CO(1-0)}$ and L$_{\rm FIR}$, as
well as between L$_{\rm FIR}$/L$^{'}_{\rm CO(1-0)}$ and $L^\prime_{\rm
CO(1-0)}$. Notice that unlike Fig.~\ref{Lfir.LCOprime-fig4.eps} , the vertical axis does not depend on
the distance. The S-R correlation analysis gives a coefficient of 0.83 and 0.59,
respectively, and both with significance level larger than 99.99\%.
Gao \& Solomon (2004b) showed that L$_{\rm FIR}$/L$^{'}_{\rm CO(1-0)}$ is
proportional to the dense molecular gas fraction as measured by L$_{\rm
HCN}$/L$^{'}_{\rm CO(1-0)}$, where HCN is the most abundant high dipole-moment
molecules. The correlations shown in
Fig.~\ref{LfirLCOprimerat.Lfir.LCOprime-fig5.eps} indicate that the dense
molecular gas fraction might increase with increasing FIR luminosities and
CO (1-0) luminosities ($L^\prime_{\rm CO(1-0)}$) for all CO detected QSOs.

\begin{figure}
\centering
\includegraphics[width=1.0\textwidth]{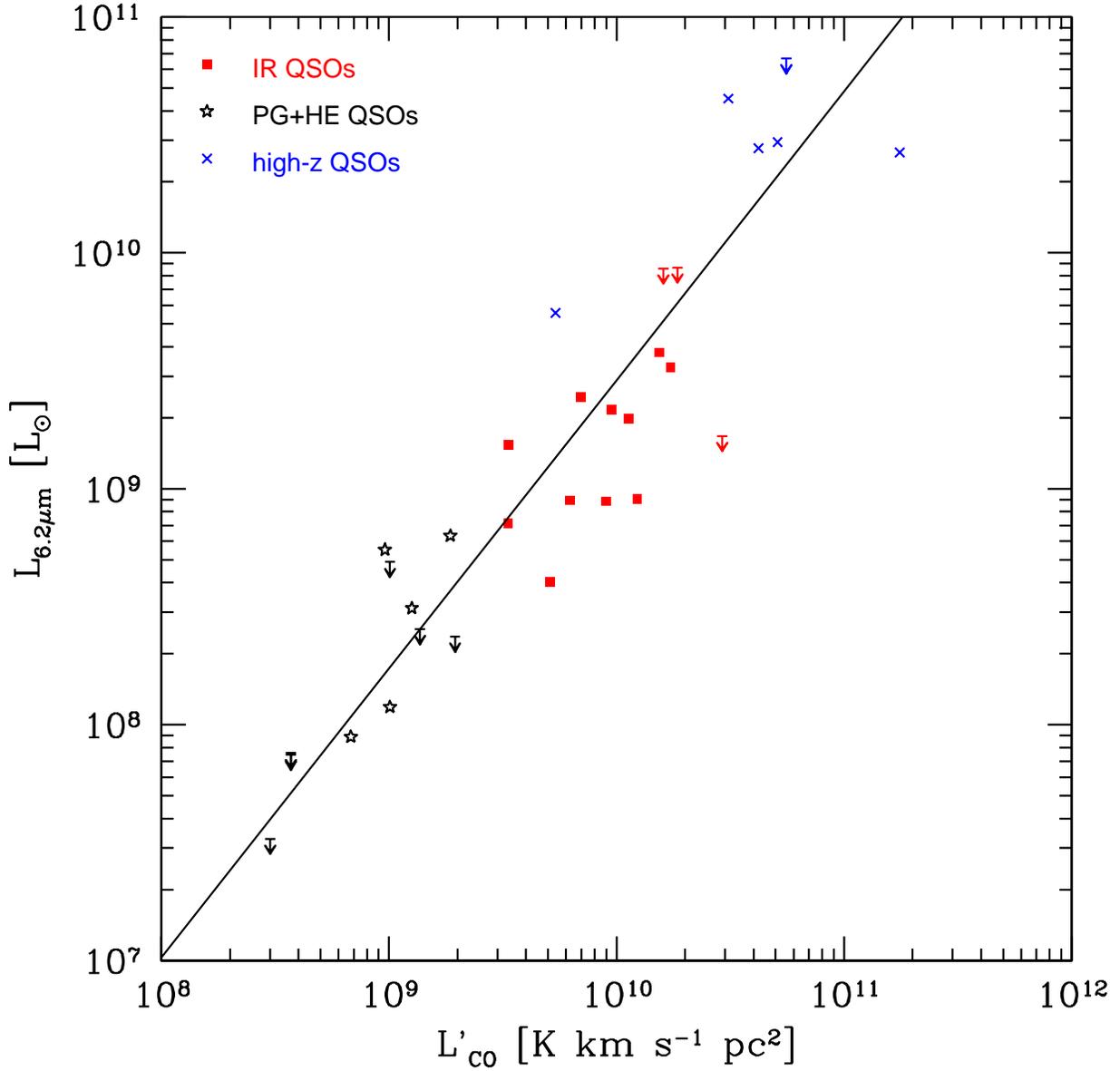}
\caption{PAH 6.2$\mu$m luminosity  L$_{\rm 6.2\mu m}$ vs. $L^\prime_{\rm CO(1-0)}$
for PAH and CO detected IR QSOs, PG$+$HE QSOs and high-redshift QSOs. The PAH luminosities of 
low-redshift and high-redshift QSOs are taken from the Quasars and ULIRGs Evolution Study
(see Schweitzer et al. 2006;  Netzer et al. 2007) and Lutz et al. (2008) respectively, whereas PAH luminosities of IR QSOs are from Cao et al. (2008).
Upper limits are plotted for some objects. A least-square bisector best-fit line is obtained for all objects, including
those with upper limits.} 
\label{L6.2um.LCOprime-fig6.eps}
\end{figure}

We also collected polycyclic aromatic hydrocarbon (PAH) 6.2$\mu$m emission data
from the IRS on {\it Spitzer} (Houck et al. 2004) for our IR QSO sample and
other CO detected QSOs (Lutz et al. 2008; Cao et al. 2008).
Fig.~\ref{L6.2um.LCOprime-fig6.eps} shows the correlation between $L_{\rm
6.2\mu m}$ and $L^\prime_{\rm CO(1-0)}$ for all sample QSOs with 6.2$\mu$m
measurements including some upper limits. Since the 6.2$\mu$m PAH emission is
believed to originate from star formation, the correlation shown in
Fig.~\ref{L6.2um.LCOprime-fig6.eps} further confirms the star formation origin
of the FIR emission for all QSOs.  Netzer et al. (2007) reached the same
conclusion based on their analysis of the SED of QSOs and ULIRGs observed by
{\it Spitzer}/IRS. Our results further strengthen the conclusion: even in the
QSO episode, the large amount of cold molecular gas reservoir still serves as the fuel
to sustain star formation seen in the FIR emission.

We now return to discuss Fig.~\ref{Lfir.LCOprime-fig4.eps}. A closer
examination of this figure reveals that the correlation between L$_{\rm FIR}$
and $L^\prime_{\rm CO(1-0)}$ for QSOs shown in
Fig.~\ref{Lfir.LCOprime-fig4.eps} is  tighter than that for all star-forming
galaxies, such as local spiral galaxies, ULIRGs and submm galaxies (SMGs, e.g.,
Fig. 8 in Bertram et al. 2007).  It appears to suggest that the cold molecular
gas properties (distribution) in QSO hosts is not as complicated as those in
starbursts, ULIRGs and/or SMGs.

This point is supported by different spatial distributions of molecular gas in
these objects.  High resolution CO observations for local IR QSOs (e.g., Mrk
231 and I Zw I, as described in \S 4.2) and high-redshift (sub)mm detected QSOs
(e.g. APM 08279+5255, Cloverleaf QSO, J1148+5251) show that the molecular gas
in these QSO hosts are concentrated in central disks/rings with radius
on the order of 1 kpc (Solomon et al. 2003; Staguhn et al. 2004; Riechers et al. 2009; Walter et al. 2009). 
Bradford et al. (2009) reached the same conclusion from analyses of many molecular line
species from the single-dish observations of the Cloverleaf.  In contrast,  a
significant fraction of ULIRGs and SMGs are interacting or merging galaxy pairs
with wide separations. In this case, a lot of the cold molecular gas has not
yet settled into nuclear rotating disks/rings, instead,  the gas is still very
extended spatially and likely has different physical parameters from the
molecular gas in the disks (Dinh-V-Trung et al. 2001; Greve et al. 2005; Tacconi et al. 2008; Riechers et
al. 2011c), which may have contributed to the larger scatters in the
correlation between L$_{\rm FIR}$ and $L^\prime_{\rm CO(1-0)}$.

Further evidence for different sizes is provided by recent high resolution
observations of SMGs and lensed high-z QSOs (Casey et al. 2011; Riechers et al.
2011c, 2011d), which reveal  extended cold molecular gas components in SMGs,
but not in lensed QSOs. The tighter correlation between L$_{\rm FIR}$ and
$L^\prime_{\rm CO(1-0)}$ for QSOs, as shown in
Fig.~\ref{Lfir.LCOprime-fig4.eps}, is consistent with different spatial
distributions of molecular gas between QSOs and ULIRGs/SMGs.

From Fig.~\ref{Lfir.LCOprime-fig4.eps}, one can also see that while most CO
detected QSOs at high-redshift are more luminous, nevertheless the local and
high-redshift CO detected QSOs form a continuous sequence. In particular, four
high-redshift CO detected QSOs have CO luminosities comparable to local IR QSOs. 
The comparable CO and FIR luminosities imply similarities between these high-redshift QSOs and
local IR QSOs in both cold molecular gas contents and SFRs. Due to sensitivity
limits of current facilities, observations of FIR and CO faint QSOs are usually
for gravitationally lensed systems: three of these four relatively faint
high-redshift QSOs are lensed (SMM 04135, IRAS F10214 and Cloverleaf).  Most
recently, Riechers (2011a) reported CO detections for four additional lensed
high-redshift QSOs with relatively low $L_{\rm FIR}$ and $L^\prime_{\rm
CO(1-0)}$ by the Combined Array for Research in Millimeter-wave Astronomy
(CARMA). Again, the intrinsic FIR and CO luminosities of these high-redshift
QSOs are comparable to our IR QSOs. The similarities can be further verified
when more observations of fainter (unlensed) high-redshift QSOs become
available with the Atacama Large (sub-) Millimeter Array (ALMA) in the near
future.

\subsection{the Evolution from Massive Starbursts to Luminous QSOs}

It is widely accepted that the growth of central SMBHs and their host
spheroids are closely related due to the local SMBH-spheroid mass relation
(Magorrain et al. 1998; Ferrarese \& Merritt 2000; Tremaine et al. 2002). This
close connection is also mirrored in similar shapes in the cosmic history of
black hole growth and star formation (e.g. Merloni et al. 2004). It
is suggestive of a possible coeval evolution of central SMBHs and their
spheroids. However, it is still unclear how or even whether spheroids of
galaxies co-evolve with their central SMBHs at all stages. In the following,
we examine more closely the relation between the accretion rate of central SMBHs
and SFRs in the host galaxies for all CO detected QSOs.

\begin{figure}
\centering
\includegraphics[width=1.0\textwidth]{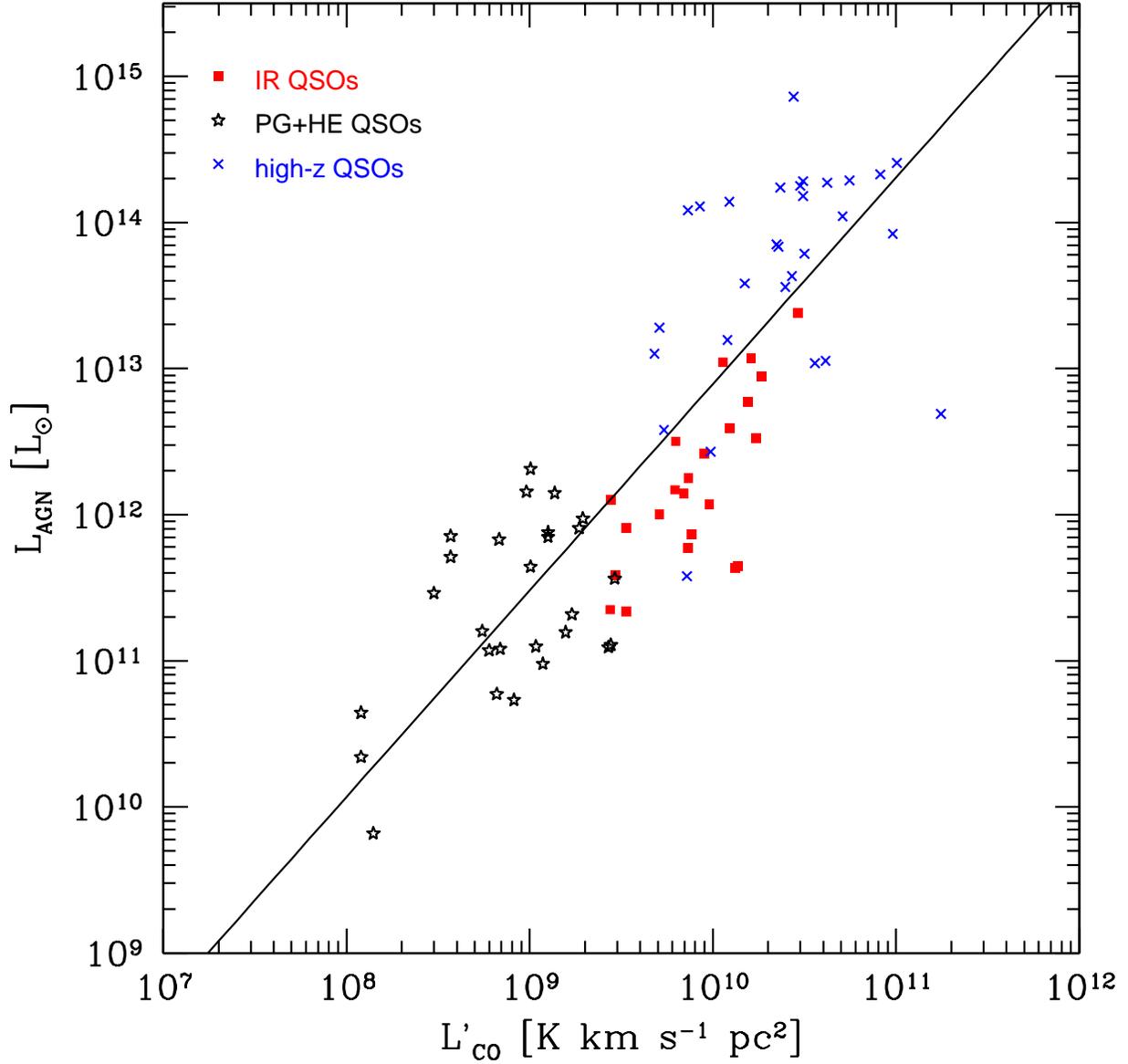}
\caption{AGN-associated bolometric luminosity ($L_{\rm AGN}$) vs. L$^{'}_{\rm CO(1-0)}$ for three samples of QSOs. 
A least-square bisector best-fit  was performed for only low-redshift CO detected PG$+$HE QSOs; the best-fit power-law slope is $1.4\pm0.2$.
}
\label{LCOprime.Lbol_fig7.eps}
\end{figure}

Fig.~\ref{LCOprime.Lbol_fig7.eps} shows the AGN-associated bolometric luminosities 
($L_{\rm AGN}$) vs. $L^\prime_{\rm CO(1-0)}$ for all CO detected QSOs.  Note that the
AGN-associated bolometric luminosities used in this paper are the integrated luminosities over
the SEDs for the AGN components, as described in \S 3.  From
Fig.~\ref{LCOprime.Lbol_fig7.eps}, we can see clearly that as $L^\prime_{\rm
CO(1-0)}$ increases, $L_{\rm AGN}$ also increases.  The Spearman Rank-order
correlation analysis gives a coefficient of 0.83 with significance $>$
99.99$\%$.  A least-square bisector fit for low-redshift CO detected PG and HE
QSOs in Fig.~\ref{LCOprime.Lbol_fig7.eps} yields a best-fit slope of $1.4\pm
0.2$.  The correlation between $L_{\rm AGN}$ and $L^\prime_{\rm CO(1-0)}$ is
suggestive of some link between the cold molecular gas on $\sim$ kpc scale
disks/rings and the central SMBH accretion process on much smaller scales - it
is likely that both nuclear star formation and SMBH accretion rely on the same
molecular gas reservoir  (Bonfield et al.  2011; Diamond-Stanic \& Rieke 2011).  

It is interesting that all IR QSOs and the few relatively faint high-redshift
QSOs with $L_{\rm AGN}$ less than $\sim 10^{13}L_\odot$ are located
on or below the best-fit line, while most high-redshift QSOs with $L_{\rm AGN}$
larger than $\sim 10^{13}L_\odot$ are located above. Such a systematic
difference in the locations on Fig.~\ref{LCOprime.Lbol_fig7.eps} for IR QSOs
and most high-redshift optically bright QSOs could reflect some intrinsic
difference in the properties between these two populations, although some
selection effects may have also played a role.

\begin{figure}
\centering
\includegraphics[width=1.0\textwidth]{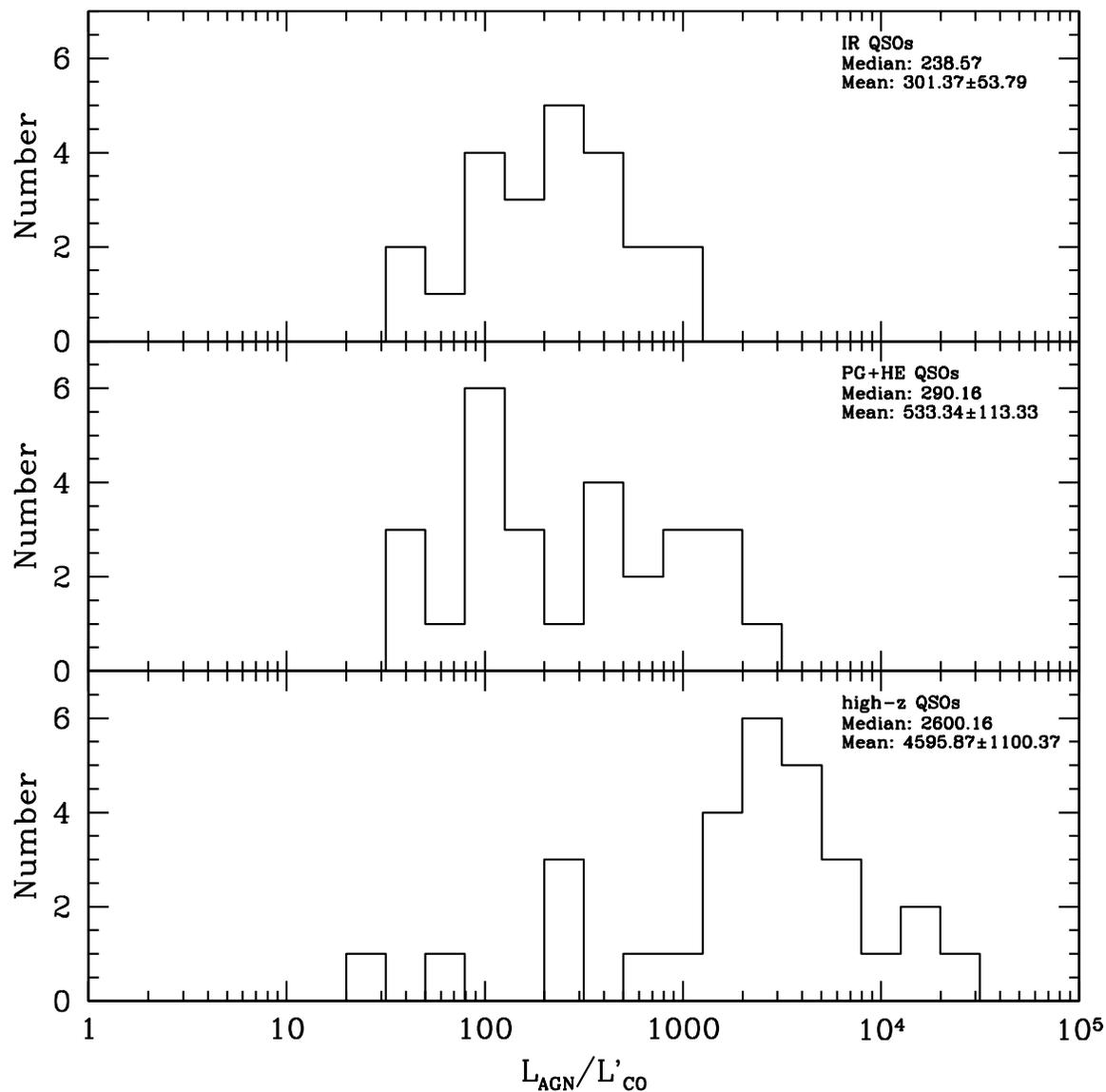}
\caption{Histogram of the $L_{\rm AGN}$ to $L^\prime_{\rm CO}$ ratio for IR QSOs (top panel), PG$+$HE QSOs (middle panel) and high-redshift QSOs (bottom panel).
The median and mean values are indicated in the top right corner of each panel. Notice that
a large fraction of optically bright PG QSOs are not detected in CO, thus the true median 
value of $L_{\rm AGN}/L^\prime_{\rm CO}$ taking into account the non-detections will be larger than that shown in the middle panel.
}
\label{bolCOrat-his-fig8.eps}
\end{figure}

The $L_{\rm AGN}/L^\prime_{\rm CO}$ ratio relates to the accretion efficiency
to central SMBH, as indicated by $\Mdot/M_{\rm gas}$.
Fig.~\ref{bolCOrat-his-fig8.eps} compares the histograms of $L_{\rm
AGN}/L^\prime_{\rm CO}$ for CO detected local IR QSOs, PG$+$HE QSOs and
high-redshift QSOs (top, middle and bottom panels respectively).  A comparison
of the top and middle panels shows that the median value of $L_{\rm
AGN}/L^\prime_{\rm CO}$ for local PG$+$HE QSOs is larger than that for IR QSOs.
Since a large fraction of optically bright PG QSOs are CO non-detected (or not
observed), the real median value of $L_{\rm AGN}/L^\prime_{\rm CO}$ for
classical QSOs should be even higher than the plotted one. The difference of
the $L_{\rm AGN}/L^\prime_{\rm CO}$ ratios between IR QSOs and the brightest
high-redshift QSOs is even more dramatic, about one order of magnitude.

\begin{figure}
\centering
\includegraphics[width=1.0\textwidth]{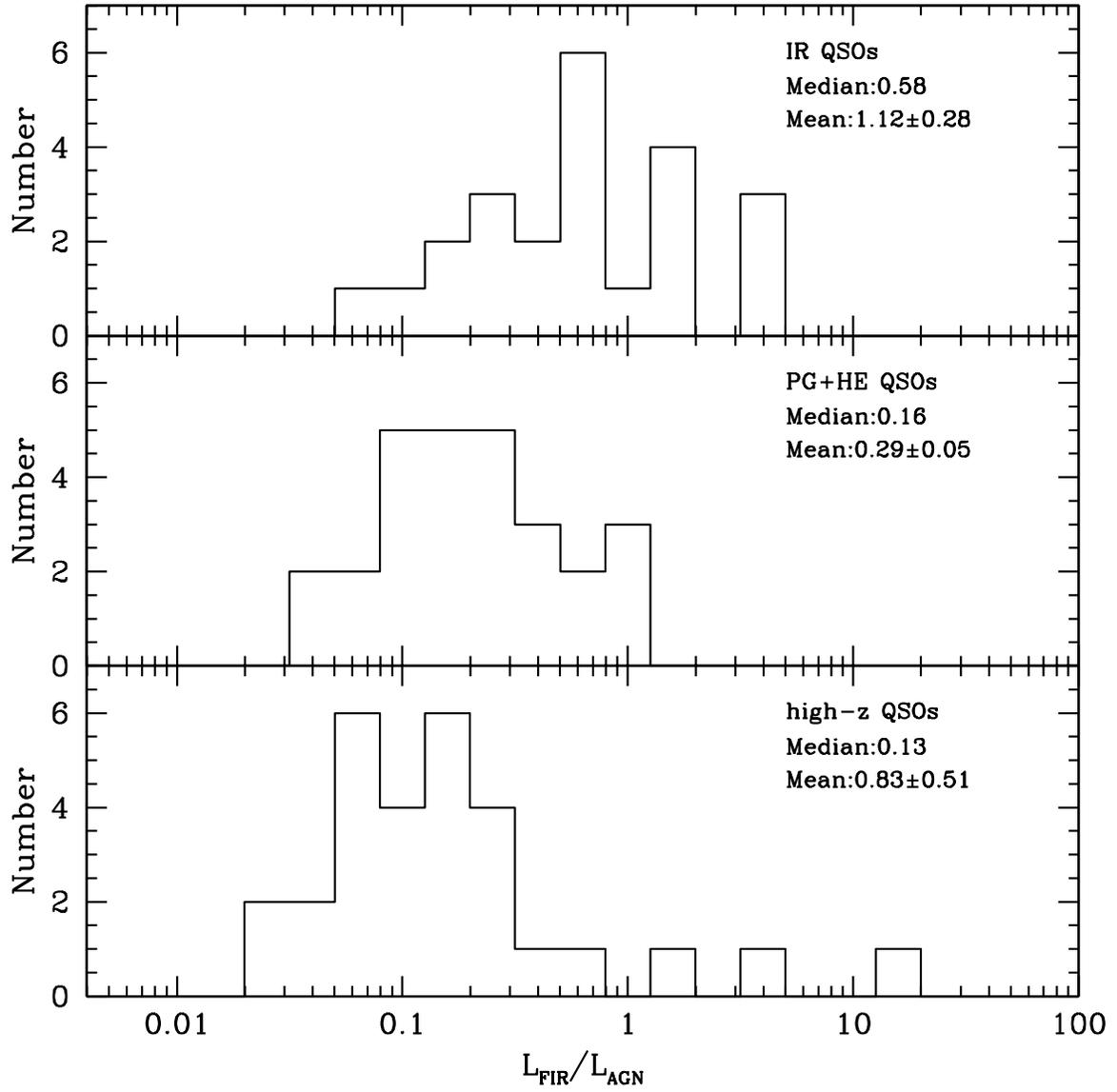}
\caption{Histogram of $\LFIR$/$L_{\rm AGN}$ for IR QSOs (top panel), PG$+$HE QSOs (middle panel) and high-redshift QSOs (bottom panel).
The median and mean values are indicated in the top right corner of each panel. 
}
\label{firbolrat.his-fig9.eps}
\end{figure}

In order to understand why the $L_{\rm AGN}/L^\prime_{\rm
CO}$ ratios for local PG$+$HE QSOs and high-redshift QSOs are much higher than
those of IR QSOs, we plot the histograms of $L_{\rm FIR}/L_{\rm AGN}$ in Fig.
~\ref{firbolrat.his-fig9.eps} for these three classes of objects.
For local classical QSOs and the brightest high-redshift QSOs, the FIR
luminosities are $\gtrsim$10\% of their central AGN bolometric luminosities,
so the main energy output for these QSOs is dominated by central AGNs,
rather than starbursts.
In contrast, the median $L_{\rm FIR}/L_{\rm AGN}$ value for IR QSOs is
$\sim$0.6, indicating that the FIR luminosities of most IR QSOs are
comparable to their central AGN bolometric luminosities.

The ratio of central SMBH accretion rate to the SFR in
their host galaxies ($\Mdot/{\rm SFR}$) can be estimated by $L_{\rm AGN}/L_{\rm FIR}$  based on the following equation:
\begin{equation}
\Mdot/{\rm SFR}=2.1L_{\rm AGN}/L_{\rm FIR}\times 10^{-3},
\label{mdoteq}
\end{equation}
assuming an accretion efficiency $\eta=0.1$ (see Hao et al.
2008). Although the uncertainties are large due to a number of assumptions
on determining these key parameters, we can nevertheless roughly
estimate whether SMBHs and their host galaxies co-evolve based on
the $L_{\rm FIR}/L_{\rm AGN}$ ratio.

As can be seen from Fig.~\ref{firbolrat.his-fig9.eps}, the median values of
$L_{\rm FIR}/L_{\rm AGN}$ are about 0.58, 0.16 and 0.13 for IR QSOs, local
PG$+$HE QSOs and high-redshift QSOs, respectively.  The corresponding median
values of $\Mdot/{\rm SFR}$ are 3.6$\times 10^{-3}$ for IR QSOs, and
1-2$\times 10^{-2}$ for local PG$+$HE QSOs and high-redshift brightest QSOs (see
also Coppin 2009).  So the $\Mdot/{\rm SFR}$ value for IR QSOs is
comparable to that of local $M_{\rm BH}/M_{\rm sph}$ value,  $\sim 1.4 \times
10^{-3}$ (H\"aring \& Rix 2004), implying their central SMBHs and host spheroids
grow in line with the local relation. In contrast, the $\Mdot/{\rm SFR}$ values
for local classical QSOs and high-redshift bright QSOs are almost an order of
magnitude larger than the local  $M_{\rm BH}/M_{\rm sph}$ value.  For these
objects, the central SMBH masses increase much faster than their host spheroids
compared to the local scaling.  It is interesting to note that the four
relatively faint high-redshift CO detected QSOs (SMM 04135, F 10214, Cloverleaf
and RXJ 124913) have high $L_{\rm FIR}/L_{\rm AGN}$ values that are close to or
even larger than those of local IR QSOs, thus the growth of their black hole may be
slower than that of high-luminosity QSOs.

According to the scenario first pointed out by Sanders et al. (1988a, b),
IR QSOs will eventually evolve toward classical QSOs. 
The different efficiencies of  gas accretion to central SMBH ($\Mdot/M_{\rm gas}$) between
the IR QSOs and local classical QSOs and the high-redshift brightest QSOs then
fit well into a picture where the QSO luminosity reaches the maximum when the
star formation at the central $\sim$ kpc decreases to a few tenths of its peak
value and no longer dominates the energy output (as indicated by the $L_{\rm
FIR}/L_{\rm AGN}$ value).  Such an evolutionary path has also been discussed
for AGNs in SMGs by Alexander (2009) and is broadly consistent with the
simulation results of Hopkins (2011).

We can also obtain more quantitative estimates of the star formation timescales
and black hole masses.  Table 3 lists the median values of SFR, $\Mdot$,
$M_{\rm BH}$, cold molecular gas mass $M_{\rm gas}$ and dynamical mass $M_{\rm
dyn}$ for IR QSOs, PG$+$HE QSOs and high-redshift QSOs.  The SFRs and accretion
rates are estimated from $L_{\rm FIR}$ and $L_{\rm AGN}$ following Hao et al.
(2008), the SMBH masses are taken from the literature,  while gas masses are
estimated from  $L^{'}_{\rm CO(1-0)}$ and dynamical masses are estimated from
the stellar CO absorption line widths in the H-band (Dasyra et al. 2006) for
local QSOs and  CO line widths for high-redshift QSOs. Such dynamical masses
should be treated as upper limits of the stellar mass of QSO's host galaxies
(e.g. Coppin et al. 2008).

First, we notice that the timescale for the gas consumption is on the scale of
a few times $10^7$ yr for all types of objects. Second, the $M_{\rm BH}/M_{\rm
dyn}$ for PG QSOs is $10^{-3}$,  similar to the value for local galaxies, again
indicating that PG QSOs follow the local scaling relation (Dasyra et al.
2006).  However, the $M_{\rm BH}/M_{\rm dyn}$ values are quite different, $4.5
\times 10^{-4}$ and $10^{-2}$ for IR QSOs and the brightest high-redshift CO
detected QSOs, respectively.  The SMBHs in IR QSOs appear to be below the local
$M_{\rm BH}$ and M$_{\rm sph}$ relation, whereas the SMBHs in the brightest
high-redshift CO detected QSOs are much larger than that predicted by the local
relation. It suggests that  while star formation and AGN activities are
intimately connected, they are not necessarily synchronized, as also suggested
by numerical simulations (Hopkins 2011).  In particular, IR QSOs appear to be
in a transition stage with a timescale of a few times $10^7$\,yr; the rapid
accretion (with $\Mdot \sim 1 M_\odot\, {\rm yr^{-1}}$) can grow the mass of a
black hole up to a few times $10^{8} M_\odot$ after this phase. Henceforth it
may follow the $M_{\rm BH}$ vs. $M_{\rm sph}$ relation seen in  local galaxies.

\section{SUMMARY}

In this paper we report the detections of cold molecular gas in 17 out of 19
infrared ultraluminous QSO (IR QSOs) hosts observed using the IRAM 30m telescope.
Including six additional IR QSOs with CO detections from the literature, our IR
QSO sample consists of 25 objects with 23 detections. The gas reservoir in these objects is found
to be several times $10^{9}M_\odot$ to a few times
$10^{10}M_\odot$. Comparisons with local ULIRGs indicate
that the cold molecular gas properties, such as the cold molecular gas mass
content, star formation efficiency, and the CO (1-0) line widths of IR QSO hosts are
similar to those of ULIRGs. These results suggest that sufficient amount of cold
molecular gas exists to sustain massive starbursts even in
the ultraluminous IR QSOs phase.

We also compared the IR QSO properties with several other QSO samples with molecular gas detections at both low
 and high redshifts.  Our main findings are summarized as follows.  
\begin{enumerate} 
\item There exists a tight correlation between L$_{\rm FIR}$ and $L^\prime_{\rm
CO(1-0)}$ for all QSOs, spanning four and three orders of magnitude in L$_{\rm
FIR}$ and $L^\prime_{\rm CO(1-0)}$, respectively.  The 6.2$\mu$m PAH luminosity
(L$_{\rm 6.2\mu m}$) and $L^\prime_{\rm CO(1-0)}$ are also closely correlated
for all sample QSOs.  
It seems to confirm the results based
on {\it Spitzer}/IRS observations that the FIR emissions of all QSOs
are mainly from star formation process rather than AGN. 
From this correlation as well as our limited CO line ratio measurements for IR QSOs 
and available CO images of QSOs, we speculate that the geometry of cold
molecular gas for most QSOs at both high-redshift and 
low-redshift could be in a disk/ring on $\sim$ kpc scale, similar to those seen in some local ULIRGs. 
\item The AGN-associated bolometric luminosities of all QSOs increase as $L^\prime_{\rm CO}$
increases, implying a possible link between the cold molecular gas on $\sim$
kpc scale (disk/ring) and the central black hole accretion process. Thus it is
likely that both star formation and central black hole accretion draw from the
same cold molecular gas reservoir.
\item The $\Mdot/{\rm SFR}$ values for IR QSOs and a few high-redshift,
relatively faint QSOs are comparable to the local M$_{\rm BH}$/M$_{\rm sph}$ value. These QSOs
might be in the transition stage from gas-rich galaxy mergers to QSOs then to
elliptical galaxies, exhibiting both high SFR and high accretion rates.
However, the local M$_{\rm BH}$/M$_{\rm sph}$ relation could not be established
in this short transition phase. If the black hole
continues to grow vigorously after this transition period, then the M$_{\rm BH}$/M$_{\rm sph}$ relation may be
established afterwards. On the other hand, for both local and very bright high-redshift QSOs, 
the black hole appears to grow much faster than the spheroids. 
It remains to be seen how the $M_{\rm BH}$ vs. $M_{\rm sph}$ evolves as a function of cosmic time.
\end{enumerate}

\acknowledgements
This work is based on observations carried out with the IRAM 30m telescope. IRAM is supported
by INSU/CNRS (France), MPG (Germany) and IGN (Spain).
We thank Drs. J. S. Huang and J. M. Wang for advice and helpful discussions.
We also thank Drs. Thomas Bertram and Lutz Wisotzki for kindly providing us with the
unpublished ${\rm B}_{\rm J}$ magnitudes for HE QSOs, and Dr. Ran Wang for
providing us data for high-redshift QSOs.
This project is supported by the NSF of China 10833006, 10973011, 11003015, 11173059 
and 973 project (2007CB815405) and Chinese Academy of Sciences (SM).

\clearpage

\clearpage
\pagestyle{empty}
\begin{deluxetable}{l c c c  c c c c}
\tablenum{1}
\tablewidth{0pt}
\tabletypesize{\tiny}
\tablecaption{CO observations of IR QSOs with the IRAM 30m telescope}                                                                 
\tablehead{ \multicolumn{1}{c}{Source} &
\multicolumn{1}{c}{R.A.} &
\multicolumn{1}{c}{Decl.} &
\multicolumn{1}{c}{$z_{\rm opt}$} &
\multicolumn{1}{c}{$z_{\rm CO}^{a}$} &
\multicolumn{1}{c}{$D_{\rm L}^{b}$} &
\multicolumn{1}{c}{Transition} &
\multicolumn{1}{c}{$t_{\rm int}^{c}$}  \\
\multicolumn{1}{c}{} &
\multicolumn{1}{c}{(J2000.0)} &
\multicolumn{1}{c}{(J2000.0)} &
\multicolumn{1}{c}{} &
\multicolumn{1}{c}{} &
\multicolumn{1}{c}{(Mpc)} &
\multicolumn{1}{c}{} &
\multicolumn{1}{c}{(min)}
}
\startdata
\multicolumn{8}{c}{Detections}\\
\tableline\\
F02065+4705  & 02 09 45.8 & +47 19 43.2 & 0.132 & 0.132 & 620 & CO(1$\to$0) & 47 \\
             &            &             &       &       &     & CO(2$\to$1) & 47 \\
F04415+1215  & 04 44 28.8 & +12 21 13.1 & 0.089 & 0.090 & 407 & CO(1$\to$0) & 85 \\
             &            &             &       &       &     & CO(2$\to$1) & 66 \\
IR06269$-$0543 & 06 29 24.7 & $-$05 45 26.0 &0.117 &0.117 & 545 & CO(1$\to$0) & 97 \\
               &            &               &       &     &     & CO(2$\to$1) & 78 \\
F07599+6508  & 08 04 30.4 & +64 59 53.3 & 0.148 & 0.149 & 703 & CO(1$\to$0) & 304 \\
             &            &             &       &       &     & CO(2$\to$1) & 219 \\
F10026+4347  & 10 05 41.8 & +43 32 41.6 & 0.178 & 0.179 & 861 & CO(1$\to$0) & 99 \\
F11119+3257  & 11 14 38.9 & +32 41 33.0 & 0.189 & 0.190 & 920 & CO(1$\to$0) & 94 \\
Z11598$-$0112  & 12 02 26.6 & $-$01 29 15.3 &0.151 &0.151 & 718 & CO(1$\to$0) &115 \\
               &            &               &      &      &     & CO(2$\to$1) &72  \\
F12265+0219  & 12 29 06.6 & +02 03 09.0 & 0.158 & 0.159 & 755  &CO(1$\to$0) &150\\  
F12134+5459  & 12 15 49.3 & +54 42 24.6 & 0.150 & 0.151  & 713  &CO(1$\to$0) &140\\ 
F13218+0552  & 13 24 19.9 & +05 37 05.0 & 0.205 & 0.204 & 1007 & CO(1$\to$0) & 113 \\
F13342+3932  & 13 36 24.0 & +39 17 32.2 & 0.179 & 0.180 & 866 & CO(1$\to$0) & 190 \\
F15069+1808  & 15 09 13.7 & +17 57 11.0 & 0.171 & 0.171 & 824 & CO(1$\to$0) & 115 \\
F15462$-$0450  & 15 48 56.8 & $-$04 59 33.5 &0.101 &0.100 & 465 & CO(1$\to$0) &143 \\
               &            &               &      &      &     & CO(2$\to$1) &91  \\
F20036$-$1547  & 20 06 31.9 & $-$15 39 05.8 &0.193 &0.192 & 942 & CO(1$\to$0) &98 \\
F21219$-$1757  & 21 24 41.6 & $-$17 44 45.3 &0.113 &0.113 & 525 & CO(1$\to$0) &198 \\
               &            &               &      &      &     & CO(2$\to$1) &137 \\
F22454$-$1744  & 22 48 04.1 & $-$17 28 28.5 &0.117 &0.118 & 545 & CO(1$\to$0) & 66 \\
               &            &               &      &      &     & CO(2$\to$1)& 66 \\
F23411+0228  & 23 43 39.7 & +02 45 05.7 & 0.091 & 0.092 & 416 & CO(1$\to$0) & 113 \\
             &            &             &             &     &  & CO(2$\to$1) & 94  \\
\tableline
\multicolumn{8}{c}{Non-detections}\\
\tableline\\
F02054+0835  & 02 08 06.8 & +08 50 05.2 & 0.345 &\nodata  & 1826 &CO(1$\to$0) &38\\
F18216+6419  & 18 21 57.3 & +64 20 36.0 & 0.297 &\nodata  & 1535 &CO(1$\to$0) &58\\
                                                                                                                                                  
\enddata
\tablecomments{Units of right ascension and declination are hours, minutes, seconds, and degrees, arcminutes, arcseconds respectively.}
\tablenotetext{a}{CO redshift was determined by computing its flux-weighted redshift, $z_{\mbox{\tiny{\rm CO}}}= \sum I(z) z/\sum I(z)$ (see, e.g., Greve et al. 2005). }
\tablenotetext{b}{$D_{\rm L}$ is the luminosity distance.}
\tablenotetext{c}{Total usable on-source integration time.}
                                                                                                                                                  
\end{deluxetable}

\clearpage                                                                                                                                                           

\begin{deluxetable}{l c c c c c c c c}
\centering
\tablenum{2}
\tablewidth{0pt}
\tabletypesize{\tiny}
\tablecaption{CO emission line and infrared properties of the IR QSO sample}
\tablehead{
\multicolumn{1}{c}{Source} &
\multicolumn{1}{c}{Transition} &
\multicolumn{1}{c}{$\Delta v_{\rm FWHM}^{a}$} &
\multicolumn{1}{c}{$\Delta v_{\rm FWZI}^{b}$} &
\multicolumn{1}{c}{$T_{\rm mb} \Delta v^{c}$} &
\multicolumn{1}{c}{$S_{\rm CO}\Delta v^{d}$} &
\multicolumn{1}{c}{$L'_{\rm CO}$$^{e}$} &
\multicolumn{1}{c}{log$L_{\rm FIR}$} &
\multicolumn{1}{c}{$L_{\rm FIR}/L'_{\rm CO}$} \\
\multicolumn{1}{c}{} &
\multicolumn{1}{c}{} &
\multicolumn{1}{c}{(km s$^{-1}$)} &
\multicolumn{1}{c}{(km s$^{-1}$)} &
\multicolumn{1}{c}{(K km s$^{-1}$)} &
\multicolumn{1}{c}{(Jy km s$^{-1}$)} &
\multicolumn{1}{c}{(10$^{9}$K km s$^{-1}$ pc$^2$)} &
\multicolumn{1}{c}{(log$L_{\odot}$)} &
\multicolumn{1}{c}{} \\
}
\startdata
\multicolumn{9}{c}{Detections}\\
\tableline\\
F02065+4705     &CO(1$\to$0) &145$\pm$12 &300 &1.84$\pm$0.17 &9.11$\pm$0.84  &7.60  &11.885 &101 \\
                &CO(2$\to$1) &172$\pm$10 &330 &5.19$\pm$0.29 &25.69$\pm$1.44 &5.36  &       &        \\
F04415+1215     &CO(1$\to$0) &112$\pm$8  &230 &1.49$\pm$0.12 &7.38$\pm$0.59  &2.75  &12.011 &373 \\
                &CO(2$\to$1) &137$\pm$16 &280 &2.48$\pm$0.29 &12.28$\pm$1.44 &1.14  &       &        \\
F06269$-$0543   &CO(1$\to$0) &258$\pm$25 &465 &2.28$\pm$0.22 &11.29$\pm$1.09  &7.36  &12.103 &172 \\
                &CO(2$\to$1) &320$\pm$22 &540 &11.00$\pm$0.68 &54.45$\pm$3.37 &8.87  &       &       \\
F07599+6508     &CO(1$\to$0) &376$\pm$14 &770 &2.36$\pm$0.12 &11.68$\pm$0.59 &12.33 &12.074 &96  \\
                &CO(2$\to$1) &486$\pm$44 &840 &7.36$\pm$0.78 &36.43$\pm$3.86 &9.60  &       &        \\
F10026+4347     &CO(1$\to$0) &343$\pm$92 &580 &0.81$\pm$0.16 &4.01$\pm$0.79  &6.20  &11.830 &109 \\
F11119+3257     &CO(1$\to$0) &285$\pm$36 &470 &1.31$\pm$0.15 &6.48$\pm$0.74  &11.33 &12.273 &165 \\
Z11598$-$0112   &CO(1$\to$0) &207$\pm$16 &580 &2.43$\pm$0.18 &12.03$\pm$0.89 &13.24 &12.260 &137 \\
                &CO(2$\to$1) &234$\pm$18 &580 &9.71$\pm$0.70 &48.06$\pm$3.46 &13.19 &       &        \\
F12265+0219$^f$     &CO(1$\to$0) &313$\pm$86 &480 & 4.88$\pm$1.41 &24.16$\pm$6.98 & 29.11 & 12.266 & 63 \\  
F12134+5459$^f$     &CO(1$\to$0) &359$\pm$84 &525 & 0.56$\pm$0.16 &2.77$\pm$0.79  &2.94   & 11.792  & 211 \\ 
F13218+0552     &CO(1$\to$0) &299$\pm$64 &430 &0.93$\pm$0.18 &4.60$\pm$0.89  &9.51 &12.175 &157 \\
F13342+3932     &CO(1$\to$0) &333$\pm$49 &520 &1.99$\pm$0.11 &9.85$\pm$0.54  &15.44 &12.125 &86  \\
F15069+1808     &CO(1$\to$0) &287$\pm$50 &440 &1.95$\pm$0.14 &9.65$\pm$0.69  &13.71 &11.875 &55  \\
F15462$-$0450   &CO(1$\to$0) &208$\pm$18 &490 &1.41$\pm$0.13 &6.98$\pm$0.64  &3.36  &11.954 &267 \\
                &CO(2$\to$1) &177$\pm$18 &370 &4.25$\pm$0.32 &21.04$\pm$1.58 &2.54  &       &        \\
F20036$-$1547   &CO(1$\to$0) &263$\pm$23 &420 &1.91$\pm$0.18 &9.45$\pm$0.89  &17.20 &12.338 &127 \\
F21219$-$1757   &CO(1$\to$0) &275$\pm$82 &400 &1.11$\pm$0.09 &5.49$\pm$0.45  &3.34  &11.631 &128 \\
                &CO(2$\to$1) &243$\pm$28 &380 &4.19$\pm$0.19 &20.74$\pm$0.94 &3.14  &       &        \\
F22454$-$1744   &CO(1$\to$0) &186$\pm$13 &380 &2.27$\pm$0.19 &11.23$\pm$0.94 &7.31  &11.548 &48 \\
                &CO(2$\to$1) &201$\pm$7  &450 &9.30$\pm$0.35 &46.04$\pm$1.73 &7.49  &       &       \\
F23411+0228     &CO(1$\to$0) &218$\pm$22 &430 &1.44$\pm$0.16 &7.13$\pm$0.79  &2.78  &11.742 &199 \\
                &CO(2$\to$1) &229$\pm$21 &390 &2.98$\pm$0.30 &14.75$\pm$1.48 &1.43  &       &        \\
\tableline
\multicolumn{9}{c}{Non-detections}\\
\tableline\\
F02054+0835     &CO(1$\to$0) &\nodata    &\nodata   &$<0.44$  &$<2.17$  &$<12.91$  &12.655  &$>350$ \\
F18216+6419$^g$     &CO(1$\to$0) &\nodata    &\nodata   &$<0.70$  &$<3.46$  &$<15.08$  &12.694  &$>328$ \\
\tableline
\multicolumn{9}{c}{Literature sources}\\
\tableline\\
PG 0050+124$^h$  &CO(1$\to$0) &370        &\nodata   &6.0$\pm$0.2  &30$\pm$1.1  &5.12  &11.417  &51 \\
Mrk 1014$^h$     &CO(1$\to$0) &270        &\nodata   &1.1$\pm$0.1  &5.5$\pm$0.5 &6.94  &12.279  &274 \\
Mrk 231$^i$      &CO(1$\to$0) &230        &\nodata   &22.0         &99.0        &8.98  &12.185  &171 \\
PG 1613+658$^h$  &CO(1$\to$0) &400        &\nodata   &1.6$\pm$0.1  &8.0$\pm$0.6 &6.23  &11.595  &63  \\
3C48$^j$         &CO(1$\to$0) &330        &\nodata   &\nodata      &1.9$\pm$0.2 &18.49 &12.666  &251  \\
PG1700+518$^k$   &CO(1$\to$0) &260        &\nodata   &0.79$\pm$0.14   &3.9$\pm$0.7 &16.07 &12.186  &95 \\

\enddata

\tablenotetext{a}{The full width at half maximum (FWHM) obtained from Gaussian fit.}
\tablenotetext{b}{The full width at zero intensity (FWZI) of the CO emission line width.}
\tablenotetext{c}{The total CO line intensity $I_{\rm CO}$ obtained by integrating $T_{\rm mb}$ over the full
velocity range, the errors were calculated using equation (1) in Matthews \& Gao (2001).} 
\tablenotetext{d}{$S_{\rm CO} \Delta v $ is the CO flux obtained using the
conversion factor S/$T_{\rm mb}$=4.95\,J$K^{-1}$}
\tablenotetext{e}{$L'_{\rm CO} = 2.45\times10^3 \left( S_{\rm CO} \Delta v
\over {\rm Jy~km~s}^{-1} \right) \left( D_{\rm L} \over {\rm Mpc}
\right) ^2 (J^{-2}(1 + z)^{-1})
[{\rm K~km~s}^{-1} {\rm~pc}^2]$  (see, e.g., Solomon et al. 1997).}
\tablenotetext{f}{Marginally detected with a signal-to-noise ratio of $\sim$3.} 
\tablenotetext{g}{Aravena et al. (2011) report the CO (1-0) detection for PG1821+643 by CARMA. The detected CO(1-0) luminosity is about one half of the 3$\sigma$ upper limit shown here.}
\tablenotetext{h}{Evans et al. (2006)}
\tablenotetext{i}{Solomon et al. (1997); Kim \& Sanders (1998)}
\tablenotetext{j}{Krips et al. (2005); Heckman et al. (1992)}
\tablenotetext{k}{Evans et al. (2009)}

\end{deluxetable}

\begin{deluxetable}{lccccccc}
\tablenum{3}
\tabletypesize{\footnotesize}
\tablecolumns{8}
\tablewidth{0pt}
\tablecaption{Median values of physical parameters for IR QSOs, PG$+$HE QSOs and high-redshift QSOs}
\tablehead{
\colhead{Name} & \colhead{$L_{\rm FIR}$}   & \colhead{SFR} & \colhead{$L_{\rm AGN}$} &
\colhead{$\dot{M}$} & \colhead{$M_{\rm BH}^{a}$} & \colhead{$M_{\rm gas}^{b}$} & \colhead{$M_{\rm dyn}^{c}$}\\
\colhead{} & \colhead{($10^{12}\, L_{\odot}$)} & \colhead{($M_\odot\, {\rm yr^{-1}}$)} &
\colhead{($10^{12}\, L_{\odot}$)} & \colhead{($M_\odot\, {\rm yr^{-1}}$)} &
\colhead{($10^{7}\, M_{\odot}$)} & \colhead{($10^{9}\, M_{\odot}$)} & \colhead{($10^{11}\, M_{\odot}$)}}
\startdata
IR QSO    &    1.2  &    391  &    1.4 &   0.9   &     5 &   6  & 0.9 \\
                                                                                                                                                         
PG$+$HE QSO    &    0.05 &   16   &    0.2 &     0.13  &    20 &   0.8 & 2\\
                                                                                                                                                         
high-redshift QSO &   7 &   2300 &      100  &        70   &    200 &   20 & 2.1 \\

\enddata
\tablecomments{The median values of parameters shown in this table are not obtained  for all our sample objects due to incomplete information for some.}
\tablenotetext{a}{The black hole masses for local IR QSOs and PG QSOs are from Hao et al. (2005); for high-redshift QSOs, 
the median value of black hole masses are taken from Coppin et al. (2008) for QSOs at z $\sim$ 2. From the most recent 
measurement by De Rosa et al. (2011), the black hole masses for QSOs at $4<z<6.5$ are $\sim 10^{9}\, M_{\odot}$, 
smaller than those at lower redshift. Notice that the median value of the black hole masses for the local classical
QSOs are only based on PG QSOs due to the lack of  $M_{\rm BH}$ information for HE QSOs. }
\tablenotetext{b}{The molecular gas mass for all QSOs is estimated from $L^\prime_{\rm CO(1-0)}$ with a conversion factor of 
$\alpha_{\rm CO}=0.8 M_\odot\, {\rm (K~km\,s^{-1}~pc^2)^{-1}}$.}
\tablenotetext{c}{The dynamical masses of IR QSOs and PG QSOs are from Dasyra et al. (2006, 2007). Notice that we assume 
the median value of dynamical mass for IR QSOs is the same as that of local ULIRGs. The dynamical masses for high-redshift QSOs
are estimated by the median value of their CO line width (see Coppin et al. 2008).}

\end{deluxetable}

\end{document}